\documentclass[aps,pra,twocolumn,groupedaddress,pdfLaTeX]{revtex4-2}

\usepackage{amsmath}
\usepackage{bm}
\usepackage{graphicx}
\usepackage{hyperref}
\hypersetup{colorlinks=true,citecolor=magenta,linkcolor=magenta,urlcolor=magenta}

\begin{document}


\title{Classical algorithm inspired by the feedback-based algorithm for quantum optimization and local counterdiabatic driving}


\author{Takuya Hatomura}
\email[]{takuya.hatomura@ntt.com}
\affiliation{NTT Basic Research Laboratories \& NTT Research Center for Theoretical Quantum Information, NTT Corporation, Kanagawa 243-0198, Japan}


\date{\today}

\begin{abstract}
We propose a quantum-inspired classical algorithm for combinatorial optimization problems, named the counterdiabaticity-assisted classical algorithm for optimization (CACAO). 
In this algorithm, a solution of a given combinatorial optimization problem is heuristically searched with classical spin dynamics based on quantum Lyapunov control of local counterdiabatic driving. 
We compare the performance of CACAO with that of quantum time-evolution algorithms, i.e., quantum annealing, the feedback-based algorithm for quantum optimization (known as FALQON), and the counterdiabatic feedback-based quantum algorithm (known as CD-FQA). 
We also study the performance of CACAO applied to large systems up to $10,000$ spins. 
\end{abstract}


\maketitle

%
%
\section{Introduction}

Combinatorial optimization was developed with various applications to operations research, e.g., assignment, transportation, searching, and scheduling~\cite{Schrijver2005}. 
Such applications can be used to resolve real-world problems. 
Recently, it has been revealed that several problems of physics~\cite{Lucas2014}, chemical and biological science~\cite{Naseri2020}, industrial problems~\cite{Yarkoni2022}, etc, can also be formulated as combinatorial optimization. 
Thus, the importance of combinatorial optimization is widely recognized in various communities. 
Quadratic unconstrained binary optimization (QUBO), also known as unconstrained binary quadratic programming (UBQP), is particularly an important class of combinatorial optimization~\cite{Kochenberger2014}. 
However, QUBO is generally NP-hard, namely, finding the exact solution of QUBO requires exponential computational time. 
The development of algorithms obtaining good approximate solutions within polynomial computational time is an important task for practical usefulness.

Solving QUBO is equivalent to the ground-state search of the Ising spin glass~\cite{Lucas2014}. 
That is, the exact solution of QUBO is given by the ground state of the Ising spin glass (the problem Hamiltonian)
\begin{equation}
\hat{H}_P=\sum_{(i,j)\in\mathcal{E}}J_{ij}\hat{Z}_i\hat{Z}_j+\sum_{i\in\mathcal{V}}h_i\hat{Z}_i,
\label{Eq.probHam}
\end{equation}
where $\mathcal{E}$ and $\mathcal{V}$ are sets of edges and vertices of a graph, and $\{J_{ij}\}$ and $\{h_i\}$ are sets of weights describing a given problem (we set $\hbar=1$ and adopt the dimensionless expression throughout the paper). 
The Pauli operators on the vertices are expressed as $\{\hat{X}_i,\hat{Y}_i,\hat{Z}_i\}$. 
The equivalence between QUBO and the Ising spin glass implies that algorithms inspired by statistical physics and/or quantum physics can be used to approximately solve QUBO.

Quantum annealing (QA) is a heuristic quantum optimization algorithm for finding the ground state or low-energy states of the Ising spin glass~\cite{Kadowaki1998}. 
In standard QA, we first prepare the ground state of a transverse-field Hamiltonian (the driver Hamiltonian) $\hat{V}=-\sum_{i\in\mathcal{V}}\Gamma_i\hat{X}_i$ with positive transverse fields $\{\Gamma_i\}$. 
By slowly changing a system Hamiltonian from the driver Hamiltonian to the problem Hamiltonian (\ref{Eq.probHam}) as $\hat{H}_\mathrm{QA}(\lambda)=\lambda\hat{H}_P+(1-\lambda)\hat{V}$ with a time-dependent parameter $\lambda=\lambda(t)$ satisfying $\lambda(0)=0$ at the initial time $t=0$ and $\lambda(T)=1$ at the final time $t=T$, we try to obtain the ground state of the problem Hamiltonian (\ref{Eq.probHam}). 
The adiabatic theorem guarantees the convergence to the ground state when the annealing time $T$ is sufficiently large~\cite{Kato1950}, while the adiabatic condition~\cite{Jansen2007} says that the exact solution cannot be obtained by QA within polynomial computational time when the energy gap closes in an exponential way.

Recently, another heuristic quantum optimization algorithm, named the feedback-based algorithm for quantum optimization (FALQON), was proposed~\cite{Magann2022,Magann2022a}. 
In this algorithm, we consider the reduction of an energy cost
\begin{equation}
E_P(t)=\langle\Psi(t)|\hat{H}_P|\Psi(t)\rangle,
\label{Eq.cost}
\end{equation}
with time evolution $|\Psi(t)\rangle$ based on quantum Lyapunov control~\cite{Cong2013}. 
As a system Hamltonian, we introduce $\hat{H}_\mathrm{FALQON}(\beta)=\hat{H}_P+\beta\hat{V}$ with a time-dependent parameter $\beta=\beta(t)$. 
By setting the parameter as $\beta(t)=i\langle\Psi(t)|[\hat{H}_P,\hat{V}]|\Psi(t)\rangle$, we can achieve the reduction of the energy cost (\ref{Eq.cost}) because the time derivative of the cost function (\ref{Eq.cost}) is given by $(d/dt)E_P(t)=-|\langle\Psi(t)|[\hat{H}_P,\hat{V}]|\Psi(t)\rangle|^2\le0$. 
This choice of the parameter $\beta(t)$ is not always the best one for the minimization of the cost function (\ref{Eq.cost}), but it guarantees the reduction of the energy until the convergence. 
Owing to the quantum nature, we cannot determine the parameter $\beta(t)$ with a single-shot process. 
In FALQON, we iteratively evaluate the parameter $\beta(t)$ by prolonging the operation time step-by-step for scheduling the total time-dependence of the system Hamiltonian $\hat{H}_\mathrm{FALQON}(\beta)$.

Shortcuts to adiabaticity are means of speeding up quantum adiabatic algorithms~\cite{Torrontegui2013,Guery-Odelin2019,Hatomura2024}. 
In counterdiabatic driving, which is a method of shortcuts to adiabaticity, we apply the counterdiabatic Hamiltonian $\hat{H}_\mathrm{CD}(t)$ to a target time-dependent Hamiltonian, and then we can completely cancel out nonadiabatic transitions~\cite{Demirplak2003,Demirplak2008,Berry2009}. 
When a target time-dependent Hamiltonian is parametrized with a time-dependent parameter as the QA Hamiltonian $\hat{H}_\mathrm{QA}(\lambda)$ with the time-dependent parameter $\lambda=\lambda(t)$, the counterdiabatic Hamiltonian is given by the following form $\hat{H}_\mathrm{CD}(t)=\dot{\lambda}\cdot\hat{\mathcal{A}}(\lambda)$, where $\dot{\lambda}=d\lambda/dt$ and $\hat{\mathcal{A}}(\lambda)$ is known as the adiabatic gauge potential~\cite{Kolodrubetz2017,Hatomura2021}. 
That is, in an impulse regime $T\to0$ ($\dot{\lambda}\to\infty$), we can ignore the target time-dependent Hamiltonian. 
Indeed, the counterdiabatic Hamiltonian $\hat{H}_\mathrm{CD}(t)$ itself has the ability to realize the adiabatic time evolution except for phase factors.

Counterdiabaticity, which is the ability of the counterdiabatic Hamiltonian or the adiabatic gauge potential to counteract diabatic changes, can also be applied to the improvement of other quantum optimization algorithms. 
The introduction of approximate counterdiabatic Hamiltonians to FALQON was discussed as the counterdiabatic feedback-based quantum algorithm (CD-FQA), and the significant improvement of the performance was reported~\cite{Malla2024}. 
In CD-FQA, we introduce $\hat{H}_\mathrm{CD\mbox{-}FQA}(\beta,\gamma)=\hat{H}_P+\beta\hat{V}+\gamma\hat{V}_\mathrm{LCD}$, where $\hat{V}_\mathrm{LCD}$ is a local approximate counterdiabatic Hamiltonian and $\gamma=\gamma(t)$ is a time-dependent parameter. 
Similarly to FALQON, the parameter setting $\beta(t)=i\langle\Psi(t)|[\hat{H}_P,\hat{V}]|\Psi(t)\rangle$ and $\gamma(t)=i\langle\Psi(t)|[\hat{H}_P,\hat{V}_\mathrm{LCD}]|\Psi(t)\rangle$ enables us to reduce the energy cost (\ref{Eq.cost}). 
More recently, the idea of digital-analog quantum computation was incorporated into CD-FQA~\cite{Chandarana2024}.

In this paper, we propose a quantum-inspired classical algorithm for combinatorial optimization, named the counterdiabaticity-assisted classical algorithm for optimization (CACAO). 
In this algorithm, the amplitude of local counterdiabatic driving is optimized by using the theory of quantum Lyapunov control.
We point out that it results in classical spin dynamics under a certain situation. 
We compare the performance of CACAO with that of QA, FALQON, and CD-FQA. 
We also study the performance of CACAO with large systems.

In Sec.~\ref{Sec.theory}, we explain the theory of CACAO. 
Numerical results are shown in Sec.~\ref{Sec.results}. 
First, we show that CACAO is not a trivial optimization process in Sec.~\ref{Sec.simple}. 
Next, we conduct a benchmark test of CACAO against QA, FALQON, and CD-FQA in Sec.~\ref{Sec.benchmark}. 
Finally, we apply CACAO to large systems in Sec.~\ref{Sec.large}. 
Section~\ref{Sec.discussion} is devoted to discussion. 
We summarize the present results in Sec.~\ref{Sec.conclusion}.

%
%
\section{\label{Sec.theory}Theory}
We derive our algorithm CACAO. 
We consider the simplest local approximate counterdiabatic Hamiltonian
\begin{equation}
\hat{H}_\mathrm{CACAO}(\{\alpha_i\})=\sum_{i\in\mathcal{V}}\alpha_i(t)\hat{Y}_i, 
\label{Eq.hamY}
\end{equation}
with time-dependent parameters $\{\alpha_i\}=\{\alpha_i(t)\}$ as a system Hamiltonian. 
This choice is based on the facts that the counterdiabatic Hamiltonian itself has the ability to realize the adiabatic time evolution and the Pauli-Y operators are the lowest-order operators in the counterdiabatic Hamiltonian for the QA Hamiltonian~\cite{Claeys2019}. 
We notice that the system Hamiltonian (\ref{Eq.hamY}) consists of the local terms for the vertices, and thus no entanglement between vertices is generated during time evolution. 
That is, when the initial state is given by a product state $|\Psi(0)\rangle=\otimes_{i\in\mathcal{V}}|\psi_i(0)\rangle$, dynamics is always given by the product state $|\Psi(t)\rangle=\otimes_{i\in\mathcal{V}}|\psi_i(t)\rangle$, where $|\psi_i(t)\rangle$ describes the state of the $i$th vertex. 
Hereafter, we assume that the initial state is given by a product state.

We determine the time-dependent coefficients $\{\alpha_i(t)\}$ based on quantum Lyapunov control. 
Similarly to FALQON and CD-FQA, we set $\alpha_i(t)=i\langle\Psi(t)|[\hat{H}_P,\hat{Y}_i]|\Psi(t)\rangle$. 
Then, it is given by
\begin{equation}
\alpha_i(t)=2\left(h_i+\sum_{\substack{j\in\mathcal{V} \\ \bm{(}(i,j)\in\mathcal{E}\bm{)}}}J_{ij}m_j^Z\right)m_i^X,
\label{Eq.alpha.COCOA}
\end{equation}
where $m_i^W=m_i^W(t)=\langle\psi_i(t)|\hat{W}_i|\psi_i(t)\rangle$ with $W=X,Y,Z$. 
Moreover, the time derivatives of $\{m_i^X(t),m_i^Z(t)\}$ are given by
\begin{equation}
\begin{aligned}
&\dot{m}_i^X=2\alpha_i(t)m_i^Z,\\
&\dot{m}_i^Z=-2\alpha_i(t)m_i^X, 
\end{aligned}
\label{Eq.COCOAeq}
\end{equation}
according to the Schr\"odinger equation.

In CACAO, we numerically solve Eq.~(\ref{Eq.COCOAeq}) with Eq.~(\ref{Eq.alpha.COCOA}). 
After the convergence or after a certain time $T$, we obtain the solution $\{m_i^Z(T)\}$ whose energy (\ref{Eq.cost}) is given by $E_P(T)=\sum_{(i,j)\in\mathcal{E}}J_{ij}m_i^Zm_j^Z+\sum_{i\in\mathcal{V}}h_im_i^Z$. 
Our algorithm is based on nonlinear classical spin dynamics, and thus it can be run on classical computers. 
When the number of the vertices is $N$ and the order of the vertices does not depend on $N$, the computational time of CACAO is given by $\mathcal{O}(NT)$. 
The worst-case computational time is $\mathcal{O}(N^2T)$, where the order of the vertices is $\mathcal{O}(N)$ and the calculation of Eq.~(\ref{Eq.alpha.COCOA}) requires an additional cost.

%
%
\section{\label{Sec.results}Results}

%
%
\subsection{\label{Sec.simple}CACAO applied to a simple system}

First, we show that CACAO is not a trivial optimization process. 
We consider a two-spin system
\begin{equation}
\hat{H}_P=J_{12}\hat{Z}_1\hat{Z}_2+h_1\hat{Z}_1+h_2\hat{Z}_2,
\label{Eq.two}
\end{equation}
as the problem Hamiltonian (\ref{Eq.probHam}). 
We assume that $J_{12}<0$, $h_1<0$, $h_2>0$, $|h_1|>h_2$, and $h_2<|J_{12}|$. 
Then, the ground state is given by the spin basis $|\uparrow\uparrow\rangle$ and the first-excited state is given by the spin basis $|\uparrow\downarrow\rangle$. 
We also assume that the initial state is given by $m_i^X(0)=1$ and $m_i^Z(0)=0$ for $i=1,2$. 
Since $m_i^Z(0)=0$, the sign of $\alpha_i(0)$ is identical with that of the longitudinal field $h_i$, but the sign of $\alpha_i(t)$ can change during time evolution. 
In numerical simulation, we fix $J_{12}=-1$ and $h_1=-1$.

\begin{figure}
\flushleft{(a) $h_2=0.9$}
\includegraphics[width=0.5\textwidth]{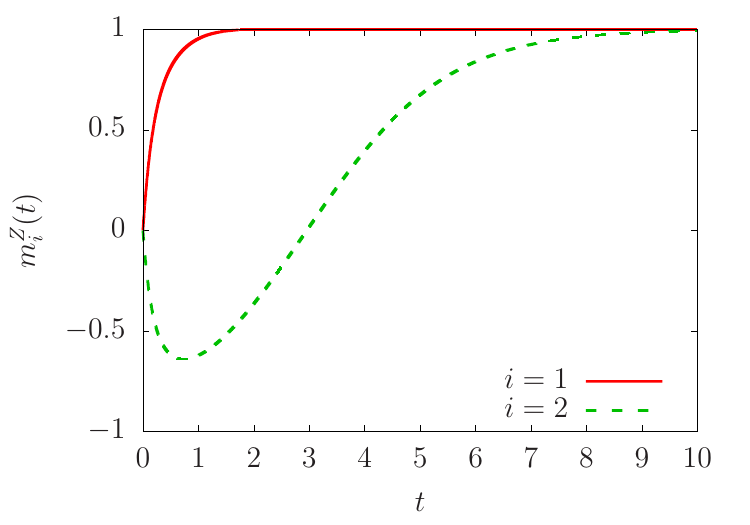}
\flushleft{(b) $h_2=0.99$}
\includegraphics[width=0.5\textwidth]{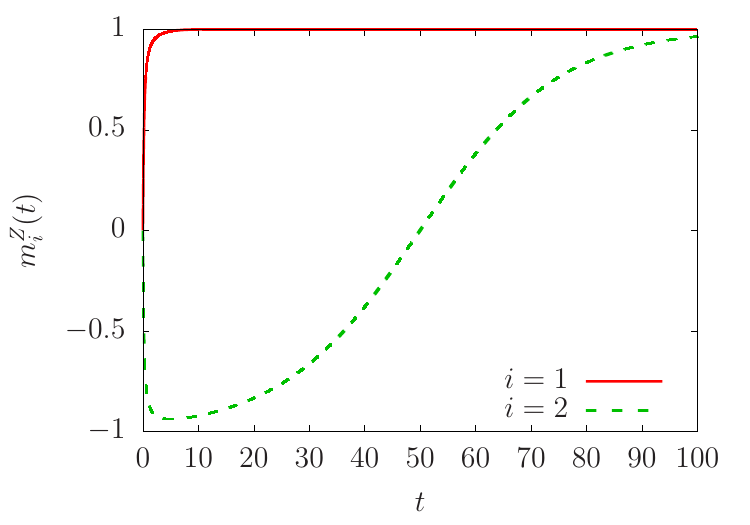}
\caption{\label{Fig.two}Time evolution of the two-spin system with CACAO. The horizontal axis is time $t$ and the vertical axis is the magnetization $m_i^Z(t)$. The red solid curve represents the spin 1 and the green dashed curve represents the spin 2. Here, (a) $h_2=0.9$ and (b) $h_2=0.99$. }
\end{figure}

We show time evolution with CACAO in Fig.~\ref{Fig.two}. 
Here, Fig.~\ref{Fig.two} (a) is that for $h_2=0.9$ and Fig.~\ref{Fig.two} (b) is that for $h_2=0.99$. 
In short timescale, the spins obey the longitudinal fields, i.e., they head for the first-excited state. 
In middle timescale, the sign of $\alpha_2(t)$ changes along with the growth of the interaction term $J_{12}m_1^Z(t)$, and thus they converge to the ground state. 
Such a nontrivial optimization process occurs even in the simple two-spin system (\ref{Eq.two}).

\begin{figure}
\includegraphics[width=0.5\textwidth]{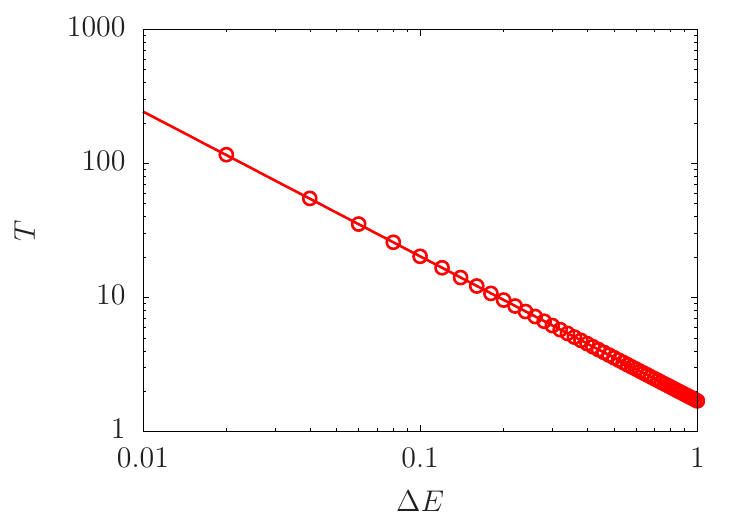}
\caption{\label{Fig.two.time}The time for convergence with respect to the energy gap of the two-spin system. The horizontal axis is the energy gap $\Delta E=-2(J_{12}+h_2)$ and the vertical axis is the time for convergence $T$, for which both the spins $m_i^Z(t)$ ($i=1,2$) are larger than 0.99. The solid line is the fitting function $T=\exp(0.5237)\Delta E^{-1.08}$. }
\end{figure}

We notice that the convergence to the ground state becomes slow as the energy gap between the ground state and the first-excited state $\Delta E=-2(J_{12}+h_2)$ becomes small [see Fig.~\ref{Fig.two} (a) and (b)]. 
Thus, we study the behavior of the time for convergence $T$, for which both the spins $m_i^Z(t)$ ($i=1,2$) are larger than 0.99, with respect to the energy gap by changing $h_2$. 
It is plotted in Fig.~\ref{Fig.two.time}. 
We find that the data clearly scale as $T=\exp(0.5237)\Delta E^{-1.08}$. 
Thus, problems become difficult for CACAO if the ground state is nearly degenerate.

%
%
\subsection{\label{Sec.benchmark}CACAO vs QA, FALQON, CD-FQA}
Next, we conduct a benchmark test of CACAO against QA, FALQON, and CD-FQA. 
As the problem Hamiltonian (\ref{Eq.probHam}), we consider the following Ising spin glass
\begin{equation}
\begin{aligned}
&\hat{H}_P=\sum_{(i,j)\in\mathcal{E}}\hat{C}_{ij},\\
&\hat{C}_{ij}=\frac{1+(-1)^{w_{ij}^{(i)}}\hat{Z}_i}{2}\frac{1+(-1)^{w_{ij}^{(j)}}\hat{Z}_j}{2}, 
\end{aligned}
\label{Eq.clause}
\end{equation}
with $w_{ij}^{(i)},w_{ij}^{(j)}=0$ or $1$. 
Here, we associate bit strings with the spin basis as $0=\uparrow$ and $1=\downarrow$, and the operator $\hat{C}_{ij}$ penalizes bit strings whose $i$th and $j$th bits are given by $w_{ij}^{(i)}$ and $w_{ij}^{(j)}$, respectively. 
The constant term does not affect dynamics and it can just be treated as a correction for the energy cost (\ref{Eq.cost}). 
Note that the ground-state search of this problem Hamiltonian is related to the 2-SAT problem~\cite{Krom1967}.

We randomly penalize bit strings whose $i$th and $j$th bits are given by ones of the following choices $(w_{ij}^{(i)},w_{ij}^{(j)})=\{(0,1),(1,0),(1,1)\}$ from the uniform distribution. 
Then, the bit string $00\dots0$ has no penalty, or in other words, the all-up spin basis $|\uparrow\uparrow\dots\uparrow\rangle$ becomes the ground state of the problem Hamiltonian (\ref{Eq.clause}) with the ground-state energy $E_\mathrm{GS}=0$.

In QA, we consider a uniform transverse field $\Gamma_i=1$ and adopt the linear schedule $\lambda(t)=t/T$ for simplicity. 
In FALQON, we also set $\Gamma_i=1$. 
In CD-FQA, we also set $\Gamma_i=1$ and use a Pauli-Y transverse-field Hamiltonian $\hat{V}_\mathrm{LCD}=\sum_{i\in\mathcal{V}}\mathrm{sgn}(h_i)\hat{Y}_i$ as the local approximate counterdiabatic Hamiltonian, where we multiply $\mathrm{sgn}(h_i)$ in analogy with variational counterdiabatic driving~\cite{Sels2017} applied to the QA Hamiltonian. 
Note that we do not take into account costs of feedback, e.g., time durations of initialization, parameter scheduling with measurement, and iterations with feedback, in numerical simulation of FALQON and CD-FQA, namely, we numerically determine the parameters with simulating single-shot processes. 
In CACAO, the initial state is set to be $m_i^X(0)=1$ and $m_i^Z(0)=0$ for all $i=1,2,\dots,N$. 
We assume that the graph $G=(\mathcal{V},\mathcal{E})$ is given by the squre lattice with the length $L$ (with the $N=L^2$ vertices) under the periodic boundary condition.

\begin{figure}
\includegraphics[width=0.5\textwidth]{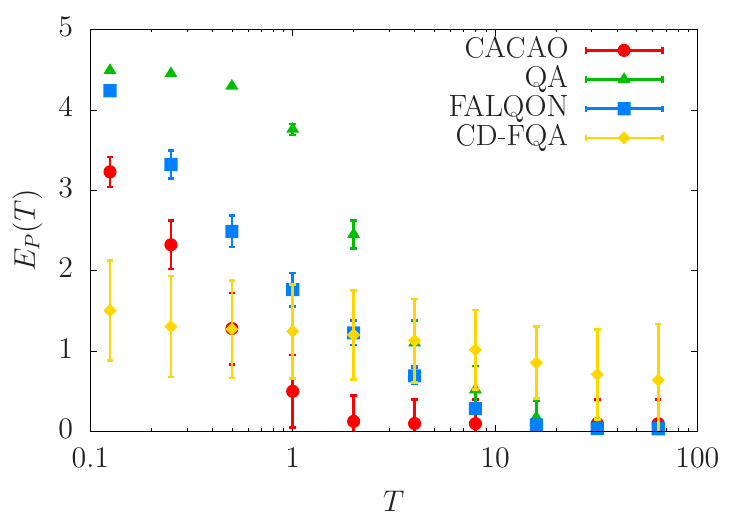}
\caption{\label{Fig.Ep_benchmark}The energy cost against the operation time. The horizontal axis is the operation time and the vertical axis is the energy cost. The red circles represent CACAO, the green triangles represent QA, the blue squares represent FALQON, and the yellow diamonds represent CD-FQA. Ten instances are sampled and the errorbars represent the standard deviation. The system size is $L=3$ ($N=9$). }
\end{figure}

We plot the energy cost (\ref{Eq.cost}) against the operation time in Fig.~\ref{Fig.Ep_benchmark}. 
Here, the system size is $L=3$ ($N=9$) and 10 instances are sampled. 
We find that CACAO shows the best convergence to low energy states, while it does not always converge to the ground state.

%
%
\subsection{\label{Sec.large}CACAO applied to large systems}

\begin{figure}
\includegraphics[width=0.5\textwidth]{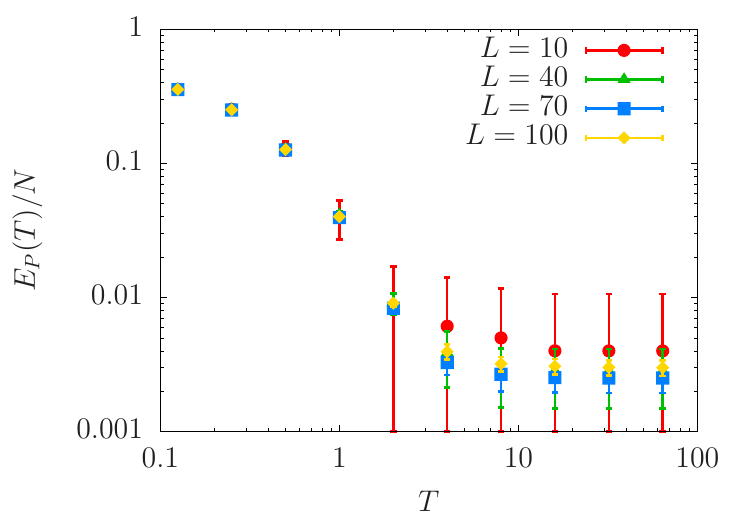}
\caption{\label{Fig.large.scale}The rescaled energy cost for CACAO against the operation time. The horizontal axis is the operation time and the vertical axis is the rescaled energy cost $E_P(t)/N$. The system size is (red circles) $L=10$ ($N=100$), (green triangles) $L=40$ ($N=1600$), (blue squares) $L=70$ ($N=4900$), and (yellow diamonds) $L=100$ ($N=10000$). Ten instances are sampled for each system size and the errorbars represent the standard deviation. }
\end{figure}

Finally, we apply CACAO to large systems. 
We consider the same problem and parameter settings as Sec.~\ref{Sec.benchmark}. 
We plot the rescaled energy cost $E_P(t)/N$ against the operation time in Fig.~\ref{Fig.large.scale}. 
Here, the system size is from $L=10$ ($N=100$) to $L=100$ ($N=10000$) and 10 instances are sampled for each system size. 
We notice that the rescaled energy costs with the various system size converge to similar values. 
On one hand, it means that finding the ground state becomes difficult for large systems since the residual energy cost is roughly proportional to the system size $N$. 
On the other hand, it means that the performance of CACAO does not change even for large systems since the energy is generally an extensive quantity. 
We also notice that the time for convergence does not depend on the system size. 
This may be the result of that the energy gap of the problem Hamiltonian (\ref{Eq.clause}) on the squre lattice does not depend on the system size.

%
%
\section{\label{Sec.discussion}Discussion}

The initial state is not necessarily $m_i^X(0)=1$ and $m_i^Z(0)=0$ in CACAO, i.e., it can be $m_i^X(0)=\cos\delta_i$ and $m_i^Z(0)=\sin\delta_i$ with $-\pi/2<\delta_i<\pi/2$. 
In particular, the parameter $\delta_i$ should be nonzero when all the longitudinal fields are zero, $h_i=0$ for all $i=1,2,\dots,N$, otherwise spins do not start to move.

Combinatorial optimization is not limited to QUBO. 
For example, we can formulate some problems as higher-order binary optimization (HUBO), and it can be reformulated as the ground-state search of spin glass models with higher-order interactions. 
We note that the theory of CACAO explained in Sec.~\ref{Sec.theory} can easily be extended to the optimization of HUBO without any theoretical advance.

There is another optimization algorithm based on classical spin dynamics assisted by counterdiabatic driving~\cite{Hatomura2018b,Hatomura2020a}. 
In that algorithm, the exact counterdiabatic Hamiltonian of a given classical spin system is used to speed up optimization. 
However, that algorithm results in divergence when classical spin dynamics meets the critical point. 
The present algorithm CACAO does not cause divergence, and thus CACAO is superior to the previous algorithm for practical use.

In numerical simulation of FALQON and CD-FQA in Sec.~\ref{Sec.benchmark}, we ignored costs of feedback, e.g., time durations of initialization, parameter scheduling with measurement, and iterations with feedback. 
However, in real setup, we have to iteratively evaluate parameters $\beta$ and $\gamma$ with prolonging the operation time step-by-step as $T=\delta t, 2\delta t,\dots,N_\mathrm{step}\delta t$ with a small time step $\delta t$ and the total number of time steps $N_\mathrm{step}$. 
Roughly speaking, for obtaining an optimized result at the target operation time $T=N_\mathrm{step}\delta t$, the actual operation time $T_\mathrm{actual}=\mathcal{O}\bm{(}\sum_{n=1}^{N_\mathrm{step}}N_\mathrm{meas}(n\delta t+T_\mathrm{ini})\bm{)}=\mathcal{O}\bm{(}N_\mathrm{step}N_\mathrm{meas}(T+T_\mathrm{ini})\bm{)}$ is necessary, where $N_\mathrm{meas}$ is the number of measurements for parameter scheduling and $T_\mathrm{ini}$ is the time duration of initialization. 
Note that further costs may be required. 
Thus, the actual performance of FALQON and CD-FQA is worse than the present numerical simulation. 
On the other hand, there is room for improvement in FALQON and CD-FQA, as well as QA. 
For example, we can introduce inhomogeneity in parameters, i.e., $\beta\to\beta_i$ and $\gamma\to\gamma_i$ for each spin. 
It may require additional costs, but the performance would be enhanced. 
Thus, we do not conclude that feedback-based quantum algorithms are worse than classical algorithms yet.

\section{\label{Sec.conclusion}Summary}

In this paper, inspired by FALQON and local counterdiabatic driving, we proposed the quantum-inspired classical optimization algorithm CACAO. 
Theoretical background of CACAO is quantum Lyapunov control of local counterdiabatic driving and the computational cost of CACAO is between $\mathcal{O}(NT)$ and $\mathcal{O}(N^2T)$, where $N$ is the number of spins and $T$ is the time for convergence or the operation time. 
We pointed out that the time for convergence might scale as $T=\mathcal{O}(\Delta E^{-1})$ with the energy gap $\Delta E$ by analyzing the two-spin system. 
We conducted the benchmark test of CACAO against QA, FALQON, and CD-FQA. 
We found that CACAO showed the best convergence to low energy states when they were applied to the spin glass model related to the 2-SAT problem. 
We also applied CACAO to the above spin glass model with large spin size and confirmed that it worked even for the system with 10,000 spins. 
We leave further benchmark tests of CACAO applied to other problems and/or against other optimization algorithms as the future work.

\begin{acknowledgments}
This work was supported by JST Moonshot R\&D Grant Number JPMJMS2061. 
\end{acknowledgments}

\bibliography{bib_CACAO}

\end{document}